# Functional State Dependence of Picosecond Protein Dynamics

J.Y. Chen[a], D.K. George[b], Yunfen He[b], J.R. Knab[c] and A. G. Markelz[b]

[a] Washington State University, Pullman, WA
[b] Physics Department, University at Buffalo, SUNY, Buffalo, NY 14260
[c] Delft University of Technology, Delft, Netherlands

## Abstract

We examine temperature dependent picosecond dynamics as a function of structure and function for lysozyme and cytochrome c using temperature dependent terahertz permittivity measurements. A double Arrhenius temperature dependence with activation energies $E_1 \sim 0.1$ kJ/mol and $E_2 \sim 10$ kJ/mol fits the native state response. The higher activation energy is consistent with the so-called protein dynamical transition associated with beta relaxations at the solvent-protein interface. The lower activation energy is consistent with correlated structural motions. When the structure is removed by denaturing the lower activation energy process is no longer present. Additionally the lower activation energy process is diminished with ligand binding, but not for changes in internal oxidation state. We suggest that the lower energy activation process is associated with collective structural motions that are no longer accessible with denaturing or binding.

Critical to the function of proteins is both their three dimensional structure and the dynamics of this structure. On the picosecond timescale motions important for function include surface solvent, side chains and large scale correlated structural motions[1-3]. It has been suggested that the large scale correlated structural motions play a key role in the necessary structural reorganization that occurs during function [4, 5]. These correlated motions have recently been measured using coherent scattering techniques [6-8]. These exciting results raise the question of whether these correlated motions change with structure and function. Further, ref. 8 has suggested evidence of an optical mode at $\sim 3$ cm$^{-1}$, suggesting that optical techniques may be used to characterize these motions.

Previously THz spectroscopic techniques have been applied to proteins [9]. To date the most dominant impact of these measurements has been the characterization of the biological water [10], and the ability of the techniques to characterize correlated motions has been disappointing. This is largely due to the dominant contribution of the biological water that likely masks the underlying protein motions.

The temperature dependence of material properties can reveal underlying processes that contribute to behavior at higher temperatures. For example, the role of conformational sub states in protein function was introduced as a result of temperature dependent measurements of ligand rebinding in myoglobin [11]. Here we find that the temperature dependent picosecond dielectric response reveals two dominant activated processes and one of these processes is removed when 3D structure is removed or constrained, suggesting that this process is associated with structural motions. The higher energy process does not significantly change with denaturing or ligand binding and is likely a beta relaxational response due to local solvent motions at the protein surface. The results demonstrate that terahertz optical measurements can access underlying correlated motions, and how these motions are effected with small ligand binding.

Previously most temperature dependent measurements of picosecond protein dynamics have focused on the so called dynamical transition (DT) at $\sim 220$K [12-14]. While not a true transition, this term is given to a rapid change in the average atomic mean square displacement $<u^2>$ as a function of temperature. The DT requires both a minimum solvent concentration and peptide size. The effect has been ascribed to a variety of phenomena, the simplest of which is the temperature dependence of beta relaxations. In addition to these local motions, collective motions will be thermally populated and contribute to the picosecond response. In order to differentiate the nature of the picosecond motions we measured the temperature dependent THz dielectric response for different functional states of two proteins, hen egg white lysozyme (HEWL) and cytochrome c (CytC).

Hen egg white lysozyme (HEWL) is a small protein, having 129 residues and a molecular weight (MW) of 14,400 Da. Lysozyme works as an antibacterial agent by hydrolysing the glycosidic bond between N-Acetyl-D-glucosamine (NAG) and N-Acetylmuramic acid (NAM) in cell walls [15]. However HEWL can stably bond with tri-N-Acetyl-D-glucosamine (3NAG) [16], and undergoes a conformational change similar to its functional state when cleaving the NAG-NAM bond.

Cytochrome *c* (CytC) is a small heme protein with 105 residues and MW of 11,700 Da. The primary role of CytC is to transfer an electron from cytochrome c reductase to cytochrome c oxidase which is embedded in the inner mitochondrial membrane, through the change in the oxidation state of the heme Fe.

We find that the temperature dependence of the imaginary part of the permittivity in the 0.2 – 2.0 THz range for native state HEWL and CytC is best described by the sum of two Arrhenius terms with activation energies $E_1 \sim 0.1$ kJ/mol and $E_2 \sim 10$ kJ/mol. With both denaturing or ligand binding the lower activation energy component is either no longer present or is strongly diminished. The denatured and bound data can be fit with a single Arrhenius term with activation energy close to $E_2$ for the native unbound state. For CytC, a double Arrhenius dependence is present for both oxidation states. When CytC is denatured, there is an increase in the permittivity and again as in the HEWL case, the lower activation energy term is no longer present. We suggest that the higher activation energy process, $E_2 \sim 10$ kJ/mol, gives rise to the DT at ~220 K seen in neutron, Mossbauer and THz measurements for a variety of peptides, proteins and nucleic acids, whereas the lower activation energy is associated with large scale collective structural motions.

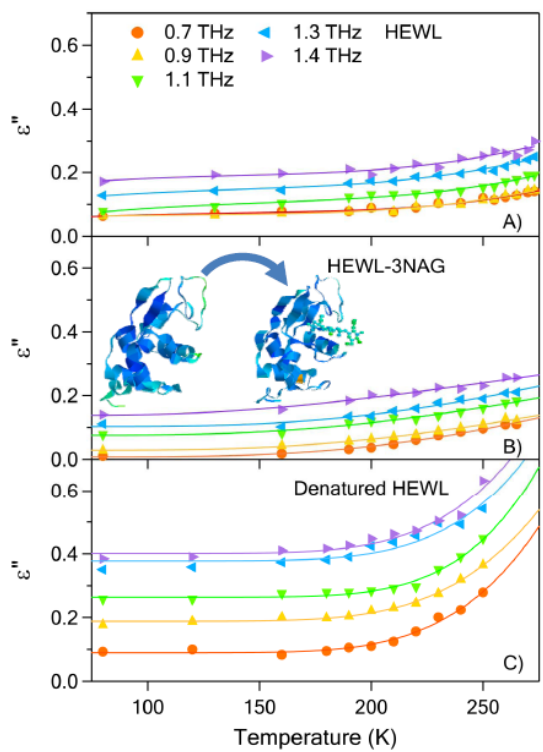

Fig. 1. Imaginary part of the permittivity versus temperature for several frequencies demonstrating systematic change with constraint and removal of structure. The same scale and legend is used for all plots. The solid lines are Arrhenius fits to the data.

Temperature dependent THz TDS was performed on solution phase samples as described earlier [17]. Starting CytC solutions were made by dissolving 30 mg of CytC powder (Bovine, Sigma C2037 or horse, Sigma C2506) into 50 µl of Tris buffer (pH 7.0). A sodium dithionite solution (60 mg/ml) was prepared to reduce the sample [18]. The oxidized CytC solution was made using 10 µl of Tris buffer and 20 µl of the starting CytC solution. The reduced CytC solution was made by adding 10 µl of the sodium dithionite solution into 20 µl of the starting CytC solution. We note that our concentrations are on the order or below those seen in the cellular environment. The oxidation state was verified by UV/Vis absorption measurements [19].

Hen egg white lysozyme (HEWL) lyophilized powder (Sigma Aldrich L6876) was dissolved in Tris buffer (pH 7.0, 0.05 M) and the final concentration of the lysozyme solution was 200 mg/ml. The HEWL+3NAG solution was made by dissolving 3NAG lyophilized powder (Toronto Research Chemicals, Inc. T735000) into the 200 mg/ml lysozyme solution with molar ratio of lysozyme to 3NAG of 1:1. The ligand binding for the solutions was verified by fluorescence measurements [20-22]. Denatured solutions (D-HEWL) were prepared by adding guanidinium hydrochloride (GdmHCl) to the native HEWL solutions to a GdmHCl concentration of 6M. Complete temperature dependence measurements of the THz permittivity for HEWL, HEWL-3NAG, DHEWL, Ferri-CytC, Ferro-CytC, and DCytC were performed on a minimum of two sample sets to verify the temperature dependences reported here.

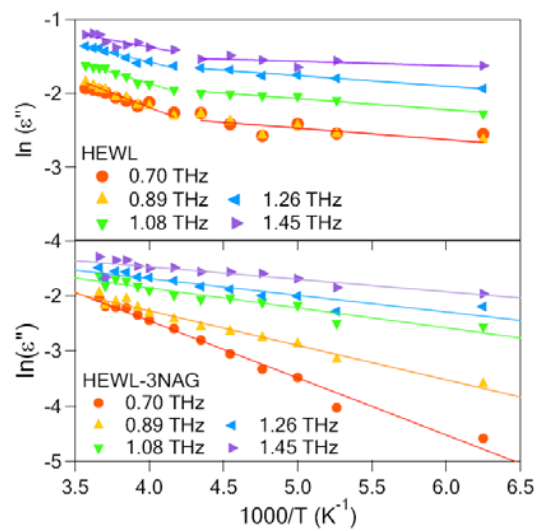

Figure 2. Arrhenius plots showing a change in the temperature dependence with ligand binding for HEWL.

In Fig. 1 we show the temperature dependent imaginary part of the THz permittivity, $\varepsilon$", for native state HEWL, 3NAG bound HEWL (HEWL-3NAG) and denatured HEWL (DHEWL). We will not discuss the real part of the permittivity $\varepsilon$' as we have previously shown that the solution phase temperature dependence of $\varepsilon$' is dominated by the bulk solvent response [23]. For HEWL $\varepsilon$" increases monotonically with temperature, with an increase in curvature above 200K. For HEWL-3NAG there is a net overall decrease in $\varepsilon$" as reported earlier [20]. The temperature dependence appears flat to 160 K then abruptly increases. This change with binding is more convincing in the Arrhenius plots in Fig. 2 which indicate a double Arrhenius dependence for native state HEWL and a single Arrhenius dependence for HEWL-3NAG.

For DHEWL $\varepsilon$" is somewhat larger than native state HEWL. Again, the temperature dependence is nearly flat to 200K and then rapidly increases. Arrhenius fits to the data are shown on Fig. 1 with a double Arrhenius fit for HEWL and a single Arrhenius fit for both HEWL-3NAG and DHEWL. A partial listing of the extracted activation energies are shown in Table 1 (full listing in supplemental documents). For HEWL we label the two extracted activation energies as $E_1$ and $E_2$, with $E_1$ referring to the lower activation energies. The extracted $E_1$ are consistent with the energies measured for collective modes [6, 7] whereas the extracted $E_2$ are consistent with those extracted from neutron quasielastic scattering and associated with the beta relaxational motions giving rise to the DT [24, 25].

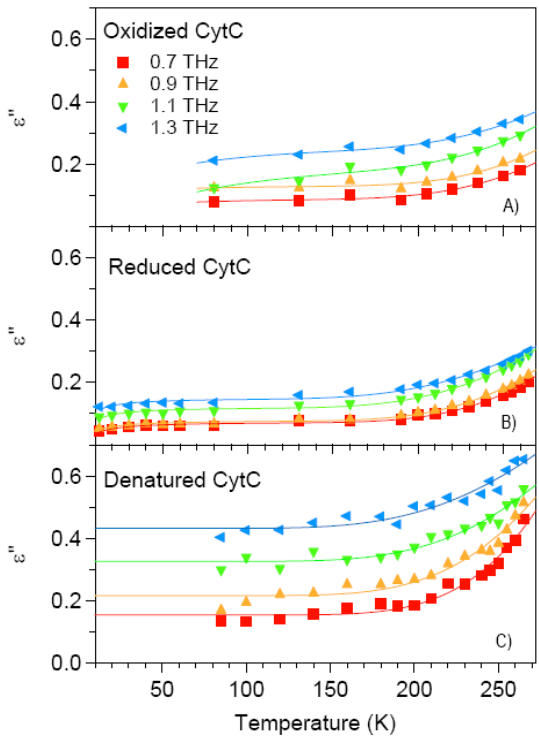

**Fig. 1 Imaginary part of the permittivity versus temperature for different oxidation states and denatured CytC. The same scale and legend is used for all plots. The solid lines are Arrhenius fits to the data.**

Fig. 3 shows $\varepsilon''$ vs. T for ferri-CytC, ferro-CytC and denatured CytC at several frequencies. $\varepsilon''$ valued for ferro-CytC are smaller than for ferri, consistent with room temperature measurements on dry CytC films [19]. As in the HEWL case, denatured CytC THz dielectric response is larger than native state for both oxidation states. For the native state CytC, there is again a monotonic increase for the entire temperature range with an increase in curvature above 200K. For DCytC, the temperature dependence is nearly flat up to 200K with a rapid increase above 200K. The temperature dependence for both ferri- and ferro-CytC are well fit with a double Arrhenius dependence as shown in the figure, whereas denatured CytC is well fit with a single Arrhenius dependence. Several of the extracted activation energies are given in Table 1. We have also found for random coil poly-lysine a good fit with a single Arrhenius dependence with activation energies similar to $E_2$ for HEWL, ferri-CytC, ferro-CytC and DCytC consistent with this activation energy arising from beta relaxation motions associated with the polypeptide surface and solvent.

We first discuss the increase in $\varepsilon''$ with denaturing for both HEWL and CytC. A number of authors have demonstrated that the water adjacent to the biomolecular surface has a larger THz permittivity than bulk[10, 26, 27]. We refer to this adjacent water as biological water. A simple explanation for the increase in the permittivity with denaturing is that the net biological water has increased with the increase in exposed molecular surface from native state to random coil. This same reasoning may be used to explain the net decrease in permittivity with 3NAG binding to HEWL. The slight decrease in exposed molecular surface with the burying of the ligand in the HEWL cleft would lead to a net decrease in biological water and subsequent decrease in net permittivity. It is also possible that correlated structural motions have been inhibited with 3NAG binding, resulting in a decrease in the dielectric response.

We now turn to the temperature dependence. Many authors have focused on the DT at 220K for hydrated biological molecules and have considered a number of possible mechanisms that give rise to that temperature dependence. This phenomenon has been somewhat contentious, in part because of the disagreement of the level of complexity that underlies it. In a glass-like system, diffusive motions are expected. Two models used to describe such systems are Vogel Tamman Fulcher (VTF) with relaxation time $\tau = \tau_o e^{(E/k_B (T-T_D))}$ and Arrhenius $\tau = \tau_o e^{(E/k_B T)}$. If one assumes that only a single temperature dependent process then VFT has been very successful in fitting the data. However, multiple temperature dependent processes may be contributing to the measured response. For example, neutron scattering measurements of $<u^2>$ reveal a linear temperature dependence at temperatures below 220K for a variety of peptides. Through the comparison of HEWL and tRNA Roh et al. [28] have found this additional linear low temperature component in $<u^2>$ arises from methyl group rotations[29].

There have been few measurements where the temperature dependence is examined as a function of binding or denaturing. In the case of HEWL the low temperature dependence changes dramatically with binding and denaturing, suggesting that the low temperature component seen in Fig. 1A) and 2A) cannot arise from methyl group rotations, as these will still be present with binding and denaturing. Furthermore methyl group rotations will not significantly contribute to our dipole coupled signal.

If we assume the simplest picture that the temperature dependence arises from two activated processes, then the data suggest that the low activation energy states are no longer accessible or are inhibited with denaturing or binding. With denaturing the structure is removed, along with the correlated structural motions. In the case of 3NAG binding, the largest amplitude mode, the hinging mode, is now constrained. Such vibrational excitations, while overdamped, will contribute to the dielectric response [30] and have been observed by other techniques. The removal of the smaller activation energy process appears to require a significant

| | HEWL | | HEWL-3NAG | DHEWL | FerroCytC | | FerriCytC | | DCytC |
|---|---|---|---|---|---|---|---|---|---|
| $\nu$ | $E_1$ | $E_2$ | $E_B$ | $E_D$ | $E_1$ | $E_2$ | $E_1$ | $E_2$ | $E_D$ |
| 0.7 | 70 | 9,722 | 7,946 | 16,485 | 63 | 14,201 | 106 | 13,684 | 14,753 |
| 0.9 | 113 | 12,118 | 7,001 | 15,657 | 52 | 13,093 | 52 | 14,713 | 12,353 |
| 1.1 | 444 | 13,411 | 6,647 | 18,781 | 40 | 11,379 | 450 | 13,708 | 11,184 |

$\nu$ in THz, $E$ in J/mol, activation energies extracted from Arrhenius fits

Table 1

structural change. For the CytC data, we see little change in the temperature dependence with oxidation. While a change in the water dynamics is expected [31, 32], there is little structural change with cytochrome c oxidation. The similarity in temperature dependence for both oxidation states of cytochrome c implies the lower activation energy is not associated with the water dynamics.

In spite of the simplicity of this model, the higher activation energies, $E_2$'s for HEWL, CytC and $E_D$ for DCytC are similar at ~ 12 kJ/mol. These energies are also consistent with those obtained from other measurements of the 220K transition such as dielectric relaxation and neutron scattering [24, 25]. $E_B$ and $E_D$ for HEWL are somewhat lower and higher respectively than the $E_2$'s. $E_1$'s are ~ 100 J/mol, or 1 meV/molecule. This is consistent with correlated motion energies found by inelastic X-ray scattering measurements on HEWL [7]. The dependence on structure and the consistency with the X-ray measurements suggests that the $E_1$ process may correspond to thermally activated large-scale correlated motions, which change with protein binding and will be present at biological temperatures.

This work demonstrates the contribution of activated protein structural motions to terahertz optical response. Further studies need to consider if the activated motion can be observed for other systems and how it changes with protein-protein interaction.

Acknowledgments: This work supported by NSF CAREER PHY-0349256.

References

[1] S. K. Pal, and A. H. Zewail, Chem. Rev. **104**, 2099 (2004).
[2] M. Brunori *et al.*, Trends Biochem. Sci. **24**, 253 (1999).
[3] B. Brooks, and M. Karplus, Proc. Natl. Acad. Sci. U. S. A. **82**, 4995 (1985).
[4] A. Rodriguez-Granillo, A. Crespo, and P. Wittung-Stafshede, J. Phys. Chem. B **114**, 1836 (2010).
[5] S. Woutersen *et al.*, Proc. Natl. Acad. Sci. U. S. A. **98**, 11254 (2001).
[6] K. Wood *et al.*, Chem. Phys. **345**, 305 (2008).
[7] D. Liu *et al.*, Phys. Rev. Lett. **101**, 135501 (2008).
[8] M. C. Rheinstadter *et al.*, Phys. Rev. Lett. **103**, 2009).
[9] A. G. Markelz, IEEE J. Sel. Topics in Quantum Electronics **14**, 180 (2008).
[10] S. Ebbinghaus *et al.*, Proc. Natl. Acad Sci. U.S.A. **104**, 20749 (2007).
[11] R. H. Austin *et al.*, Biochemistry **14**, 5355 (1975).
[12] S. Khodadadi *et al.*, Biochim. Biophys. Acta **1804**, 15 (2010).
[13] W. Doster *et al.*, Phys. Rev. Lett. **104**, 098101 (2010).
[14] G. Chen *et al.*, Philosophical Magazine **88**, 3877 (2008).
[15] D. L. Nelson, and M. M. Cox, *Lehninger Principles of Biochemistry* (Worth Publishers, New York, 2000), Third edn.
[16] D. C. Phillips, Proc Natl Acad Sci U S A. **57**, 483 (1967).
[17] Y. He *et al.*, Phys. Rev. Lett. **101**, 178103 (2008).
[18] K. A. Schenkman *et al.*, J. Appl. Physiol. **82**, 86 (1997).
[19] J.-Y. Chen *et al.*, Phys. Rev. E. Rapid **72**, 040901 (2005).
[20] J.-Y. Chen *et al.*, Appl. Phys. Lett. **90**, 243901 (2007).
[21] S. S. Lehrer, and G. D. Fasman, J. Biol. Chem. **242**, 4644 (1967).
[22] Y. Amo, and I. Karube, Biosensors and Bioelectronics **12**, 953 (1997).
[23] A. G. Markelz *et al.*, Chem. Phys. Lett. **442**, 413 (2007).
[24] W. Doster, and M. Settles, in *Hydration Processes in Biology: Theoretical and Experimental Approaches*, edited by M.-C. Bellissent-Funel (IOS Press, Amsterdam, 1999).
[25] E. Cornicchi *et al.*, Biophys. J. **91**, 289 (2006).
[26] S. J. Kim *et al.*, Angewandte Chemie International Edition **47**, 6486 (2008).
[27] U. Heugen *et al.*, Proc Natl Acad Sci U.S.A. **103**, 12301 (2006).
[28] J. H. Roh *et al.*, Biophys. J. **96**, 2755 (2009).
[29] J. H. Roh *et al.*, Biophys. J. **91**, 2573 (2006).
[30] J. Knab, J.-Y. Chen, and A. Markelz, Biophys. J. **90**, 2576 (2006).
[31] R. Gnanasekaran, Y. Xu, and D. M. Leitner, J. Phys. Chem. B **114**, 16989 (2010).
[32] Y. F. He *et al.*, Biophys. J. **100**, 1058 (2011).

# Functional State Dependence of Picosecond Protein Dynamics: Supplemental Material

J.Y. Chen[a], D.K. George[b], Yunfen He[b], J.R. Knab[c] and A. G. Markelz[b]
[a]Washington State University, Pullman, WA
[b] Physics Department, University at Buffalo, SUNY, Buffalo, NY 14260
[c] Delft University of Technology, Delft, Netherlands

|  | HEWL | | HEWL-3NAG | Denatured HEWL |
| --- | --- | --- | --- | --- |
| ν (THz) | $E_1$(J/mol) | $E_2$(J/mol) | $E_B$(J/mol) | $E_D$(J/mol) |
| 0.7 | 70 | 9,722 | 7,946 | 16,485 |
| 0.9 | 113 | 12,118 | 7,001 | 15,657 |
| 1.1 | 444 | 13,411 | 6,647 | 18,781 |
| 1.3 | 225 | 11,794 | 7,416 | 16,962 |
| 1.4 | 137 | 12,564 | 5,062 | 16,114 |

Table 1: Activation energies extracted from the temperature dependent ε" for HEWL.

|  | Reduced CytC | | Oxidized CytC | | Denatured CytC |
| --- | --- | --- | --- | --- | --- |
| ν (THz) | $E_1$(J/mol) | $E_2$(J/mol) | $E_1$(J/mol) | $E_2$(J/mol) | $E_D$(J/mol) |
| 0.5 | 85 | 15,653 | 416 | 12,943 | 14,753 |
| 0.7 | 63 | 14,201 | 106 | 13,684 | 12,353 |
| 0.9 | 52 | 13,093 | 52 | 14,713 | 11,184 |
| 1.1 | 40 | 11,379 | 450 | 13,708 | 9,845 |
| 1.3 | 31 | 10,014 | 180 | 12,787 |  |
| 1.6 | 12 | 7,267 | 260 | 17,351 |  |

Table 2: Activation energies extracted from the temperature dependent ε" for cytochrome c.